\documentclass[twocolumn, APS,showpacs, prl]{revtex4}
\usepackage{graphicx}
\usepackage{dcolumn}
\usepackage{bm}
\usepackage{wrapfig}
\usepackage{amssymb}
\def\be{\begin{equation}}
\def\ee{\end{equation}}
\def\e#1{\label{#1}\end{equation}}
\def\bea{\begin{eqnarray}}
\def\eea{\end{eqnarray}}
\def\ea#1{\label{#1}\end{eqnarray}}

\def\bem#1{\begin{mathletters}\label{#1}}
\def\eml{\end{mathletters}}

\def\ket#1{{|#1\rangle}}
\def\bra#1{{\langle#1|}}

\def\4#1{{\boldsymbol{#1}}}
\def\8#1{{\widetilde{#1}}}
\def\bse{\begin{subequations}}
\def\ese{\end{subequations}}

\def\Rb87{$^{87}\text{Rb}$}
\def\0{\ket{0}}
\def\1{\ket{1}}

\begin{document}
\title{Dark Entangled Steady States of Interacting Rydberg Atoms}
\author{D. D. Bhaktavatsala Rao}
\author{Klaus M{\o}lmer}
\affiliation{%
Department of Physics and Astronomy, Aarhus University, Ny Munkegade 120, DK 8000 Aarhus C,
Denmark. \\
}%
\date{\today}
\begin{abstract}
We propose a scheme for rapid generation of high fidelity steady state entanglement between a pair of atoms. A two-photon excitation process towards long-lived Rydberg states with finite pairwise interaction, a dark state interference effect in the individual atoms, and spontaneous emission from their short-lived excited states lead to rapid, dissipative formation of an entangled steady state. We show that for a wide range of physical parameters, this entangled state is formed on a time scale given by the strengths of coherent Raman and Rabi fields applied to the atoms, while it is only weakly dependent on the Rydberg interaction strength.
 \end{abstract}
 \pacs{42.50.Gy, 42.50.Dv, 03.67.-a, 42.50.-p}
\maketitle

The strong blockade interaction between Rydberg excited atoms open  many possibilities to explore neutral atoms for quantum computing and for the study of a variety of complex many-body and light-matter problems \cite{revmod}. The first proposal by Jaksch et al.\cite{jaks} to use Rydberg blockade to implement a fast two-qubit controlled-NOT (CNOT) gate has been followed by a variety of schemes for fast quantum gates with atomic ensembles \cite{luk, saff1,saff2,zheng}, entangled state preparation \cite{klaus1}, quantum algorithms \cite{chen,klaus2}, quantum simulators \cite{weimer}, and efficient quantum repeaters \cite{exp4}. Fidelities of around $0.9$ for performing a $CNOT$ gate (probability truth table) and $0.7$ for generation of entanglement between two atoms using the Rydberg blockade interaction have been reported in \cite{exp1,exp2,exp3}. These fidelities are mainly limited by the finite magnitude of the blockade interaction with respect to the exciting lasers and errors due to spontaneous emission from  the intermediate state used in the two-photon coupling to the Rydberg state \cite{zhang}. To reduce these errors one must excite very high lying Rydberg states with large blockade interactions and one must apply excitation fields with a large intermediate state detuning.
\begin{figure}
\includegraphics[width=80mm,height=60mm]{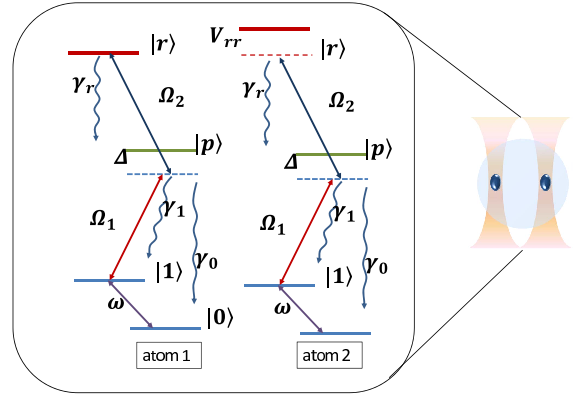}
\label{setup}
\caption{Schematic representation of two atoms driven by laser and Raman fields $(\Omega_1,\Omega_2,\omega)$ and coupled via Rydberg state interaction $V_{rr}$. The atoms are described as four-level systems comprised of two ground levels $\ket{1}$ and $\ket{0}$, coupled weakly to each other by a Raman process, while the ground level $\ket{1}$ is strongly coupled to the Rydberg level $\ket{r}$ by a resonant two-photon process via the intermediate short-lived excited level $\ket{p}$. }
\end{figure}
Dissipation provides an attractive supplement to unitary interactions for the preparation of few-atom entangled states and can even be tailored to implement universal quantum computation \cite{weimer,vers,diehl}. This has drawn further attention to the use of dissipation as an active ingredient in quantum information processing. In this Letter, we show that decay by spontaneous emission of light can be used as a key resource to generate pairwise entangled steady states of atoms with interacting Rydberg states. We shall show that our scheme works for even moderate interactions between the excited atoms and that the convergence to the entangled steady state is rapid enough to yield robustness against realistic noise and loss mechanisms.

The physical setup for our dissipative generation of entangled states involves two atoms with two ground hyperfine states, labeled $\ket{0}$ and $\ket{1}$ \cite{levels}, and a Rydberg state $\ket{r}$, which can be excited via the intermediate state $\ket{p}$, see Fig. 1. The atoms are trapped within a distance of few tens of $\mu$m, such that they experience a non-zero energy shift $V_{rr}$, when both atoms occupy the Rydberg state $\ket{r}$. We apply resonant excitation from  the state $\ket{1}$ to the optically excited state $\ket{p}$ and from $\ket{p}$ to the Rydberg level $\ket{r}$ with Rabi frequencies $\Omega_1$ and $\Omega_2$, respectively, and we drive the transition between states $\ket{0}$ and $\ket{1}$,  by a resonant Raman process with strength $\omega$. A detuning $\Delta$ may be applied with respect to the intermediate level $\ket{p}$, but in the following we will set $\Delta=0$. The Rydberg and optically excited states decay by spontaneous emission of radiation. We assume that the Rydberg state lifetime is much longer than the optical state lifetime $\gamma_R << \gamma_p$.

To briefly describe the main idea behind our entanglement mechanism, it is useful to consider first the states of a single atom in the absence of the Raman coupling field. The laser fields give rise to the electromagnetically induced transparency (EIT) phenomenon \cite{harris} associated with the coherent trapping of atomic population in the dark eigenstate $\ket{D} = \frac{1}{\Omega}[{\Omega_2\ket{1}-\Omega_1\ket{r}}]$, where $\Omega=\sqrt{\Omega_1^2+\Omega_2^2}$. The atom may also reside in the uncoupled ground state $\ket{0}$, and if we apply a weak resonant Raman coupling between states $\ket{0}$ and $\ket{1}$ with coupling strength $\omega$ ($\omega < \Omega_1,\Omega_2$), it is a good approximation to replace the bare state $\ket{0}\rightarrow\ket{1}$ coupling by an effective coupling between the eigenstates, $\ket{0}$ and $\ket{D}$, under the strong Rabi field interaction. This coupling has the strength $\omega\Omega_2/\Omega$ given by the state $\ket{1}$ component of $\ket{D}$.

Under the same approximation, the dark singlet combination,
\begin{equation} \label{darksinglet}
\ket{DS} = \frac{1}{\sqrt{2}}\left(\ket{D0}-\ket{0D}\right),
\end{equation}
of two atoms is invariant under the application of the strong Rabi and the weak Raman coupling. The triplet space of states $(\ket{00},\ (\ket{D0}+\ket{0D})/\sqrt{2},\ \ket{DD})$ are also dark states, \textit{i.e.}, they do not couple to the short lived excited atomic state, but the Raman coupling causes rotations among them. In the absence of interatomic interaction, any initial product state of the atoms, will on the time scale of a few $\gamma_p^{-1}$ evolve into an incoherent mixture of the dark singlet and triplet states with no entanglement.

Since the singlet state contains only one atom in the state $\ket{D}$ and hence only single Rydberg state components, it is immune to the interaction term $V_{rr}$, while the triplet component $\ket{D D}$, due to the interaction, is energetically shifted away from the ground state manifold and seizes to be an eigenstate of the interaction. This destroys the EIT mechanism \cite{cadams,petro,sevilay} and $|DD\rangle$ now couples to the rapidly decaying states. As a consequence of this coupling, also the other two triplet states, which are coupled by the Raman field, acquire finite lifetime and undergo excitation and decay by spontaneous emission until the atomic population accumulates in the dark singlet state (\ref{darksinglet}).  Ideally, when the blockade strength becomes infinite, all the population will rapidly return into states with maximally one atom in the state $\ket{D}$. The presence of finite blockade has the same effect, but may require a longer time $(\sim 4\Omega^2/\Omega^2_1V_{rr})$ to remove also the population from the doubly excited long lived Rydberg states. When the associated rate is larger than the Raman field strength $\omega$, the system does not distinguish between finite and infinite blockade strength, and  contradictory to the condition for unitarily generated entanglement, which strictly requires $V_{rr} \gg \Omega$, the much weaker condition
\be
\label{cond}
 V_{rr}>\left(\frac{2\Omega}{\Omega_1}\right)^2\omega,
\ee
will ensure the formation of a high fidelity entangled state. This entangled steady state is obtained from any initial state of the atoms, assuming neither a very large nor a particularly precisely defined value of the interaction.

For a quantitative analysis of the dissipative preparation scheme, we shall solve the two-atom master equation numerically and we shall investigate and interpret the detailed dependence of our results on the physical parameters of the problem.
\begin{figure}
\begin{center}
\includegraphics[width=90mm]{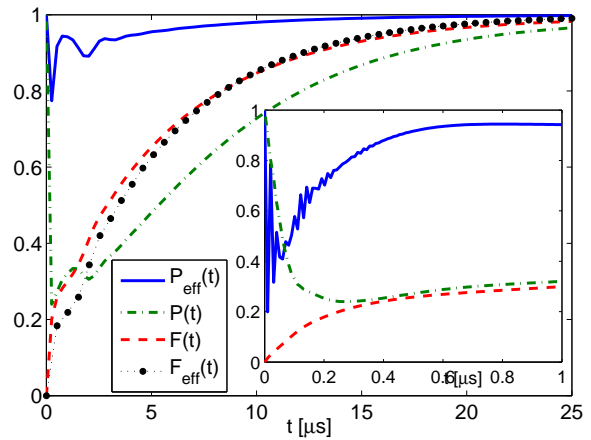}
\end{center}
\label{setup1}
\caption{(Color online) The fidelity $F(t)=\bra{DS}\rho(t)\ket{DS}$ (red dashed line), the purity of the two-atom state $P(t)$ (green dot-dashed line) and the population of the effective four-level subspace $P_{eff}(t)$ (blue solid line), are plotted as functions time. Shown in the inset is the early time variation of the same quantities. An analytical estimate for the fidelity $F_{eff}(t)$ (black dotted) obtained from the effective four-level subspace is also shown in the figure. The parameters chosen for the calculation are $\Omega_1/ 2\pi = 20$ MHz, $\Omega_2 = 2\Omega_1$, $\omega/2\pi = 250$kHz, $\gamma_1/2\pi=\gamma_2/2\pi = 3.03$ MHz, $V_{rr} = \Omega_1$ and $\gamma_R/2\pi=1$kHz.}
\end{figure}
The total Hamiltonian describing the interacting Rydberg atoms is
\be
{H} = H_1\otimes\mathcal{I}+\mathcal{I}\otimes H_2 + V_{rr}\ket{rr}\bra{rr},
\ee
with ($j=1,2$) single atom Hamiltonian operators,
\bea
\label{ham}
H_{j} = \omega\ket{1}_{jj}\bra{0}+\Omega_1\ket{1}_{jj}\bra{p}+\Omega_2\ket{p}_{jj}\bra{r}+h.c,
\eea
where we assume a resonant coupling to take full advantage of the rapidly decaying intermediate level $\ket{p}_j$.

The evolution of the system due to spontaneous emission is described by the master equation
\be
\partial_t \rho = i[\mathcal{H}_{eff},\rho] + \sum_{j,k} \mathcal{C}^{{(j)}^\dagger}_k\rho\mathcal{C}^{(j)}_k, ~~\mathcal{H}_{eff} = \mathcal{H}-\frac{i}{2}\sum_{j,k}\mathcal{C}^{{(j)}^\dagger}_k\mathcal{C}^{(j)}_k
\ee
where $\mathcal{C}^{(j)}_k$ are Lindblad operators, which describe the decay processes of the $j-th$ atom by spontaneous emission of light, $\mathcal{C}^{(j)}_0 = \sqrt{\gamma_0}\ket{0}_{jj}\bra{p}, ~\mathcal{C}^{(j)}_1 = \sqrt{\gamma_1}\ket{1}_{jj}\bra{p}$ ($\gamma_p=\gamma_0 +\gamma_1$). Rydberg state decay is described by similar terms, \textit{e.g.}, $\mathcal{C}^{(j)}_R = \sqrt{\gamma_r}\ket{p}_{jj}\bra{r}$, but since the Rydberg state is long lived, we defer our discussion of its consequences for the scheme to the end of the Letter.

\begin{figure}
\begin{center}
\includegraphics[width=90mm]{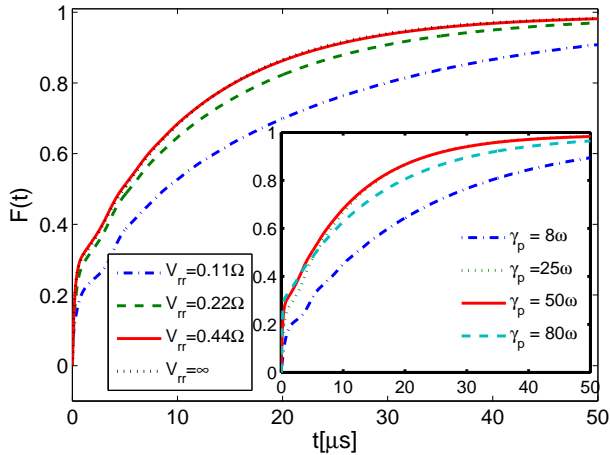}
\end{center}
\label{setup1}
\caption{(Color online) The dependence of the fidelity $F(t)$ on time is shown for four different values of $V_{rr}/\Omega$ keeping $\gamma_p$ fixed. In the inset is shown the variation of $F(t)$ for different values of the decay rate $\gamma_p$ for perfect blockade condition. For the above simulation the other fixed parameters $\Omega_1/ 2\pi = 20$ MHz,$\Omega_2 = 2\Omega_1$, $\omega/2\pi = 125 $kHz, $\gamma_1/2\pi=\gamma_2/2\pi = 3.03$ MHz and $\gamma_R/2\pi=1$kHz.}
\end{figure}
In Fig. 2 we have plotted the fidelity $F(t)=\bra{DS}\rho(t)\ket{DS}$, and the purity $P(t)=$Tr($\rho(t)^2$) of the state determined by solution of the two-atom master equation, starting from the product state $\ket{11}$ (the results are similar for any other initial state).
The results confirm the evolution described above, with an initial rapid evolution towards an entangled state fidelity of approximately $0.25$ and a comparable low value of the purity, compatible with the atoms incoherently populating product states of $\ket{0}$ and $\ket{D}$ (see the inset for a magnified view of the initial dynamics).

To explain the results in Fig. 2 we shall describe the long time behavior within the effective four-level subspace spanned by the basis vectors $\ket{T_0} = \ket{DD},~\ket{T_1} = \frac{1}{\sqrt{2}}[\ket{D0}+\ket{0D}],~\ket{T_2} = \ket{00},~\ket{T_3}\equiv \ket{DS} = \frac{1}{\sqrt{2}}[\ket{D0}-\ket{0D}]$. The effective unitary evolution can be found by projecting the total Hamiltonian (\ref{ham}) on to this subspace, which gives
\bea
\mathcal{H} = \tilde{\omega}\left(\ket{T_0}\bra{T_1} + \ket{T_1}\bra{T_2}+h.c.\right),~\tilde{\omega}\sim\sqrt{2}\omega\left(\frac{\Omega_2}{\Omega}\right).
\eea
In addition to the pure unitary evolution there is a strong coupling of the level $\ket{T_0}$ to the high energy spectrum. The rapid decay from these high energy states effectively appear as if the population in $\ket{T_0}$ is decaying back into all four states of the subspace, and the effective decay rate $\Gamma$ will be determined by the Rabi fields, $\gamma_p$, and $V_{rr}$. The new Lindblad operators (assuming perfect blockade) are given by
\be
{C}_j = \sqrt{\frac{\Gamma}{4}}\ket{T_j}\bra{T_0},~\Gamma \sim \gamma_p\left(\frac{\Omega_1}{\sqrt{2}\Omega}\right)
\ee
where $j=0,~1,~2,~3$. As $\ket{T_3}$ is a zero eigenvalue state of the Hamiltonian and a dark state with respect to the dissipative dynamics, it accumulates the population on a time scale determined by $\tilde{\omega}$ and $\Gamma$. In Fig.2 one can see an excellent agreement between the exact numerical simulations and the dynamics predicted by the effective four-level dynamics described here. For the parameters used in the plot, $\gamma_p/2\pi = 6.07$MHz, $\Omega_2/2\pi=2\Omega_1/2\pi = 40$MHz and $\omega/2\pi = 0.25$MHz, the effective decay rate and frequency are  $\Gamma/2\pi = 1.89$MHz, and $\tilde{\omega}/2\pi = 320$kHz. To further confirm the effective four level dynamics we have plotted the population of the effective four level subspace $P_{eff}(t)$ obtained from the full master equation analysis. One can see that all the population resides in the four level subspace for $t > 1/\Gamma$. Hereafter, on a longer time scale the system reaches a steady state with a high overlap with the dark singlet state $\ket{DS}$, (\ref{darksinglet}).

As long as the coupling and interaction parameters fulfil $\Omega_1,~\Omega_2,~\gamma, V_{rr} \gg \omega$ and $\Omega_2 > \Omega_1$, Fig. 3 reveals a very weak dependence of the fidelity $F(t)$ on the dissipative atomic decay rate $\gamma_p$ and the interaction strength $V_{rr}$ (for more information see \cite{suppl}). From the same figure one can see that for $V_{rr}/\Omega \ge 1/4$, and $\gamma_p/\omega \ge 20$ ($\gamma_p/\Omega \ge 0.1$), the formation of the entangled state is almost purely determined by the interplay of dynamics between the Rabi and Raman fields. This is surprising since spontaneous decay and Rydberg interactions are both essential physical components for the formation of the entangled state. Only when either of these strengths are small in comparison with the Rabi fields, the convergence to steady state becomes slower.
In addition to the deviations from these conditions the fidelity also gets reduced by other long time effects such as decay of the Rydberg state which reduces the steady state fidelity $F(\infty) \approx 1-\gamma_R/\Gamma$ \cite{suppl}, and dephasing due to {e.g,} magnetic noise.
The bounds on the steady state fidelity and the rate of entanglement can be determined from the eigenspectrum of $\mathcal{H}_{eff}$ (5). The maximum fidelity achievable is given by the overlap of the state $\ket{DS}$ with the eigenstate of $H_{eff}$ which has an eigenvalue with the smallest imaginary part \cite{suppl}. As the imaginary part of the energy determines the decay rate for the corresponding eigenstate the convergence towards the state $\ket{DS}$ is determined by the gap between the eigenvalues with lowest imaginary component \cite{suppl}.

The steady state (1) is a maximally entangled state of two atomic qubits with states $\ket{0}$ and $\ket{D}$, and as it requires only finite Rydberg interaction, it may be prepared with atoms quite far apart, e.g,, within a regular neutral atom array. Such an entangled state may be used to teleport qubits and thus to perform long distance gates in a neutral atom quantum computer \cite{chuang}. To accommodate local operations one may use the Rydberg blockade gate mechanism, and since that assumes atomic qubits encoded initially in the bare atomic ground states, we may need to turn $\ket{DS}$ in (1) into a state of the pure ground state form $\frac{1}{\sqrt{2}}(\ket{10}-\ket{01})$. Note that this can be done by adiabatically turning off the $\Omega_1$ field, or by abruptly switching the phase of the lasers to drive the $\ket{D}$ superposition state into the atomic state $\ket{1}$.

In summary, we have presented a fast and robust scheme for dissipative generation of entangled steady states of a pair of atoms with interacting Rydberg states. The scheme does not demand the interaction to be in the blockade regime, and provided an experimentally natural hierarchy between the magnitudes of the different coupling terms, it does not depend on fine tuning of any parameters. Note that the role of $V_{rr}$ is merely to perturb the state $\ket{DD}$ so that the triplet states are no longer dark states of the system. For realistic parameters used in the simulations we see that any product state evolves into the desired entangled state within few tens of microseconds. This implies that, under steady driving conditions, our scheme protects the desired entangled state by automatically compensating for the harmful effects of decoherence and decay.

The present analysis also shows that the EIT dark state feature, which is lost in the presence of interaction for the case of two three level atoms \cite{cadams}, is regained by the weak coupling to an additional ground state. The effective $\ket{0}$ and $\ket{D}$ two-level description of the atoms leads to the identification of singlet and triplet-like states of two atoms, and we believe it may be a good starting point to study correlated effects under similar conditions in larger number of Rydberg interacting atoms, e.g. along the lines of \cite{Saffmandiss}.

\begin{acknowledgements}
       The authors thank Mark Saffman for helpful comments and suggestions. This work was supported by the project MALICIA under FET-Open grant number $265522$, and the IARPA MQCO program.
\end{acknowledgements}

\section{Supplementary Information}

\noindent
\subsubsection{Atomic level schemes}
While a non-degenerate stable ground state and metastable excited state may, indeed, be found in alkaline earth atoms \cite{lemke,lowedyck,jensen} , our description in the main text of the qubit states  $\ket{0}$ and $\ket{1}$ as non-degenerate eigenstates of a bare atom Hamiltonian is not in accord with the level scheme of most neutral atom candidates for quantum computing and information purposes. Our analysis can, however, be applied for, e.g., rubidium atoms in one of the following, slightly modified, forms : (i) The states $\ket{0(1)}$ and $\ket{p}$ may be implemented as the “stretched” hyperfine ground state with $F=1, M=1$,  $(F=2, M=2)$, and excited state with $F’=3,M=3$ driven by circularly polarized fields, so that the transitions are closed, even under spontaneous decay \cite{weid}. The $\ket{p}$ state then decays only to $\ket{1}$  and not to $\ket{0}$,  as we allow in the main text, but since any transiently populated state converges to the dark state, this will only affect details in this transient evolution. The single atom stretched states are field sensitive, but their singlet combination is decoherence free under the influence of field fluctuations which are homogeneous over the separation between the atoms. (ii) The states $\ket{0}$ and $\ket{1}$ may be implemented as the magnetic field insensitive ground states with   $F=1, M=0$, and $F=2, M=0$, respectively, and the atoms may be excited with linearly polarized light. A constant magnetic field on the atoms will shift the Zeeman ground states with $M$ different from zero. A realistic shift of a few $MHz$ suffices to detune the two-photon resonant transitions and hence prevent the dark state mechanism for excitation ladders with $M\ne 0$. As a consequence, atoms decaying into the corresponding Zeeman ground states will become excited and undergo decay, until they eventually reach the $M=0$ dark state, which remains the only dark state of the system. Simulations with multi-level systems verify that this mechanism works, also in the presence of additional fields serving to repump population from unwanted components of the $\ket{0}$  state manifold.

\noindent
\subsubsection{Entangled state fidelity}
We have shown in the main text that after a nonexponential transient behavior at short time $t<1/\gamma_p$, the fidelity $F(t)$, converges towards unity in an exponential manner,
\be
F(t>1/\gamma_p) = F_{max}(1-d{\rm e}^{-\gamma_{E}(t-\tau)}),
\ee
where $1-d = \bra{DS}\rho(\tau)\ket{DS}$ is the population in $\ket{DS}$ after the initial transient time $\tau$, $\gamma_E$ is the rate of formation of the entangled state, and $F_{max}$ is the maximum attainable fidelity for given physical parameters$(\Omega_1, \Omega_2, \omega, \gamma_p, V_{RR})$.

\noindent
\subsubsection{Rate of entanglement formation}
In Fig. 1(Left) we show the variation of $\gamma_E$ as a function of $\omega$ keeping all other parameters fixed. The results shown are obtained by numerical solution of the master equation (Eq. 5, main text), and one clearly sees a predominantly linear dependence on small $\omega$, while a deviation is seen for larger values of $\omega$. The calculations also show that if $\gamma_p$ is increased proportionally with $\omega$ the linear behavior holds even for large $\omega$.

In Fig. 1(Right) we show the variation of $\gamma_E$ as a function of $\Omega_1$, keeping the other parameters fixed. One can see that $\gamma_E$ depends on the ratio $\epsilon = \Omega_2/\Omega_1$, while it is independent of the absolute magnitudes of the Rabi-frequencies $\Omega_1$ and $\Omega_2$. Using these observations and assuming values for $\gamma_p$ and $V_{RR}$ in the saturation regime (see Fig. 3a of main text), the rate of entanglement formation can be approximated by
\be
\gamma_{E} \sim \omega\left(\frac{\Omega_1}{4\Omega}\right)g(\gamma,V_{RR}),
\ee
where the function $g(\gamma,V_{RR})$ is a constant of order unity for a wide range of parameters of the model.
\begin{figure*}
\includegraphics[width=80mm]{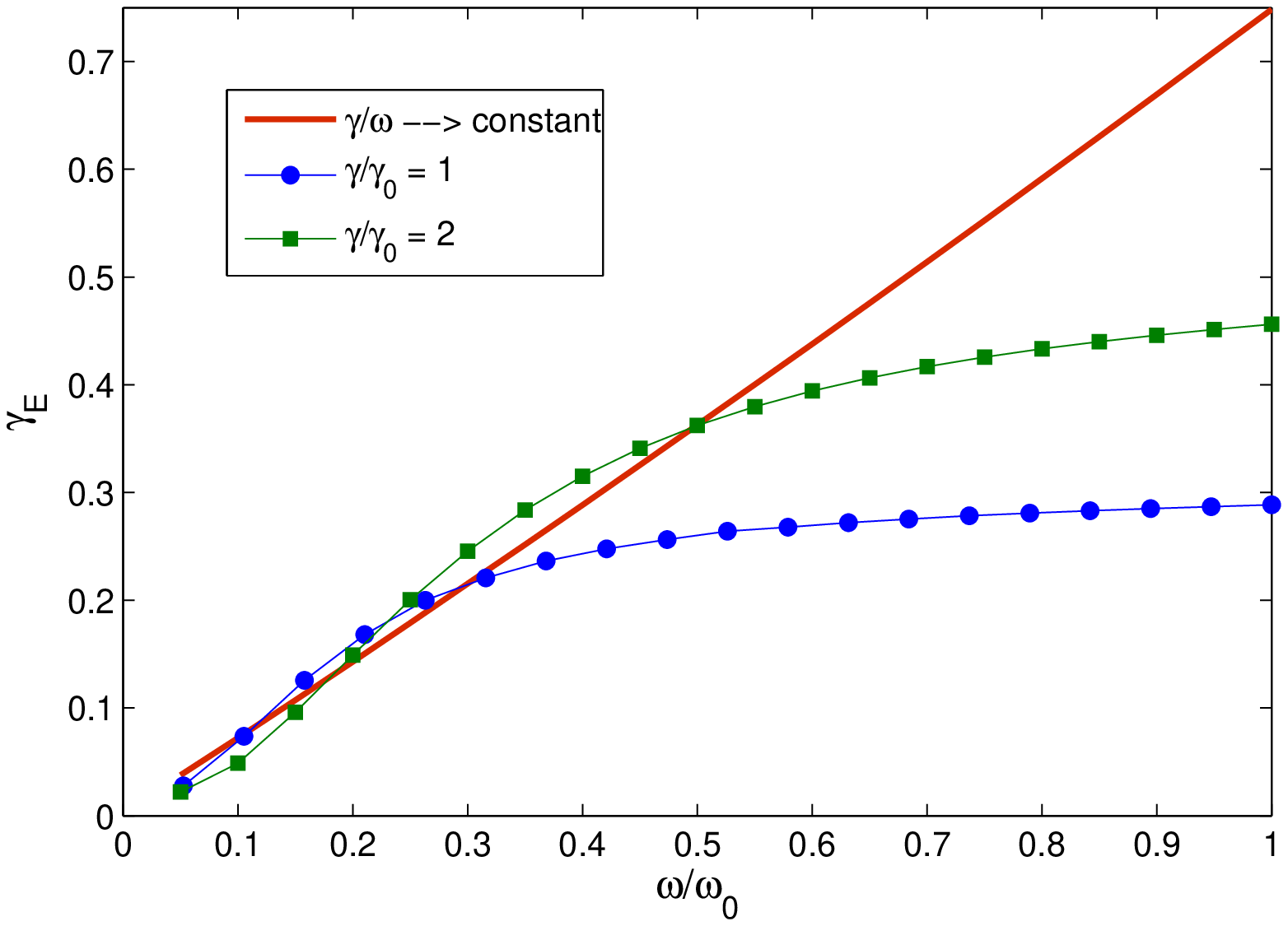}
\includegraphics[width=70mm]{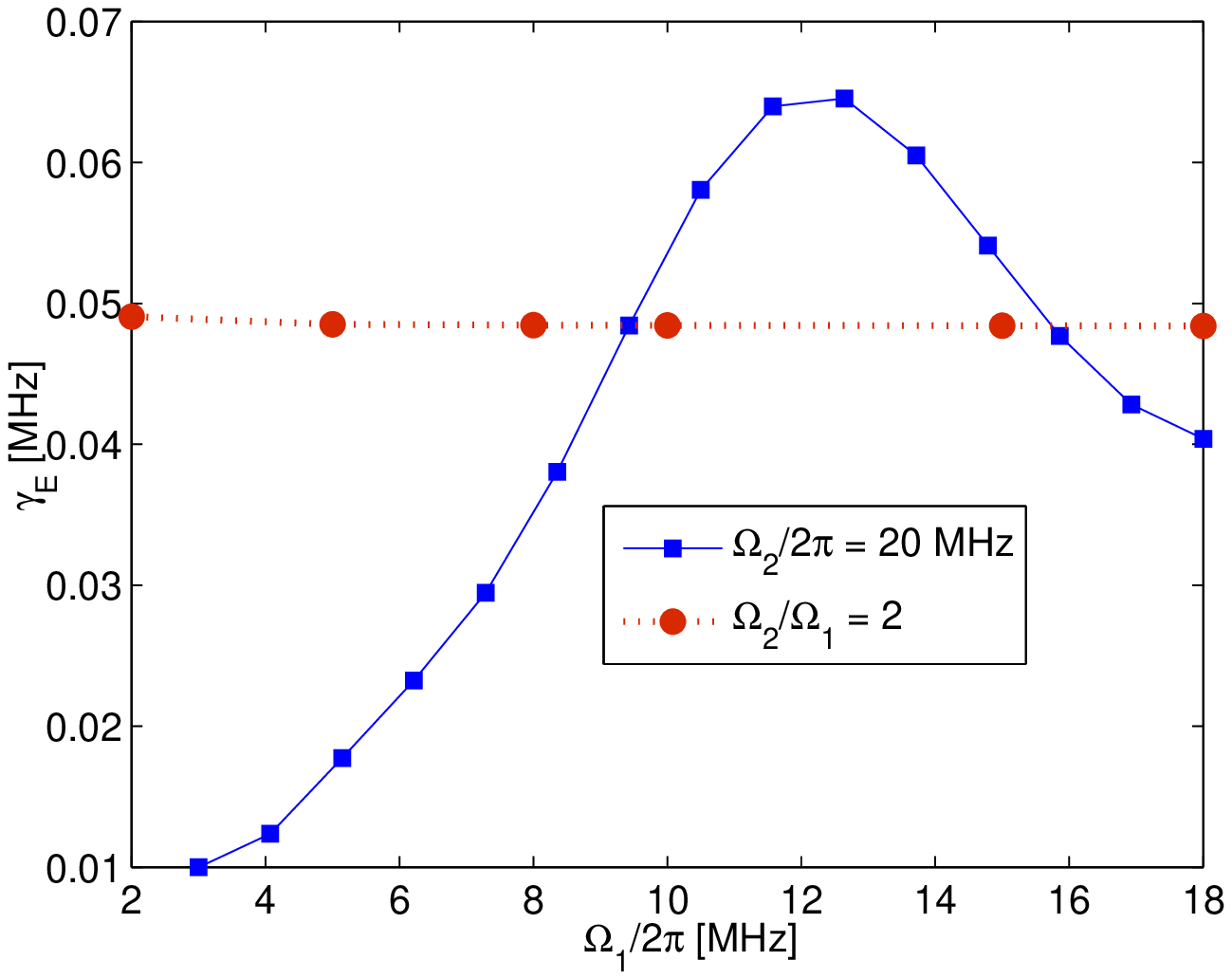}
\caption{(Left) The entanglement rate $\gamma_E$ is plotted as a function of the Raman frequency $\omega$, for three different cases (as shown in the legend). Other parameters used in the simulation are $\Omega_1/2\pi = 10MHz, ~\Omega_2 = 2\Omega_1, ~V_{RR} = \Omega_1$ and $\omega_0/2\pi = 1$MHz, $\gamma_0 = 6$MHz. (Right) The entanglement rate is plotted as a function of the Rabi frequency $\Omega_1$ for two different cases: (i) $\Omega_2$ is kept fixed (blue squares) and (ii) $\Omega_2$ is varied keeping $\Omega_2/\Omega_1$ fixed (red circles). For this plot, $\gamma_p/2\pi =1$MHz, $\omega/2\pi = 0.1$MHz.}
\end{figure*}

\noindent
\subsubsection{Maximum fidelity}
We now analyze the asymptotic fidelity achievable with the proposed scheme. In the case of non-interacting atoms, $\Omega_1$ determines the rate at which population accumulates in the effective two-level subspace spanned by the single-atom dark state $|D\rangle$ and $|0\rangle$. If $\Omega_2 \gg \Omega_1$, the ground state component of the dark state increases but at the same time the convergence rate to the entangled steady state decreases. If, conversely, $\Omega_1 > \Omega_2$, the dissipative processes suppress the sensitivity to the Rydberg interaction $V_{RR}$, mediated by $\Omega_2$ on the $\ket{p}\rightarrow\ket{r}$ transition, and $\Omega_2 > \Omega_1$ is a quite stringent requirement to achieve the high fidelity singlet state. In Fig 2 (Left) we present numerically calculated values of $F(t)$ and $F_{max}$ for different values of the Rabi frequencies. As shown in the inset, the singlet state fidelity is governed by $\epsilon$, with the optimal choice for both the convergence rate and for the steady state fidelity given by $1.7 \le \epsilon \le 2.3$.

In Fig. 2 (Right) we show the dependence of $F_{max}$ on the Raman coupling $\omega$ and the Rydberg state life time $\gamma_R$. Clearly, there exists an optimal regime in the parameter space to achieve fidelities close to unity. Though a finite $V_{RR}$ is enough to produce the state $\ket{DS}$, a too small value of $V_{RR} \ll \Omega$ (see Fig. 3a main text) will slow down the entangling process. This would in turn increase the errors due the finite lifetime of the Rydberg state $\ket{r}$, leading to a decrease in the value of $F_{max}$.

\begin{figure*}
\includegraphics[width=70mm]{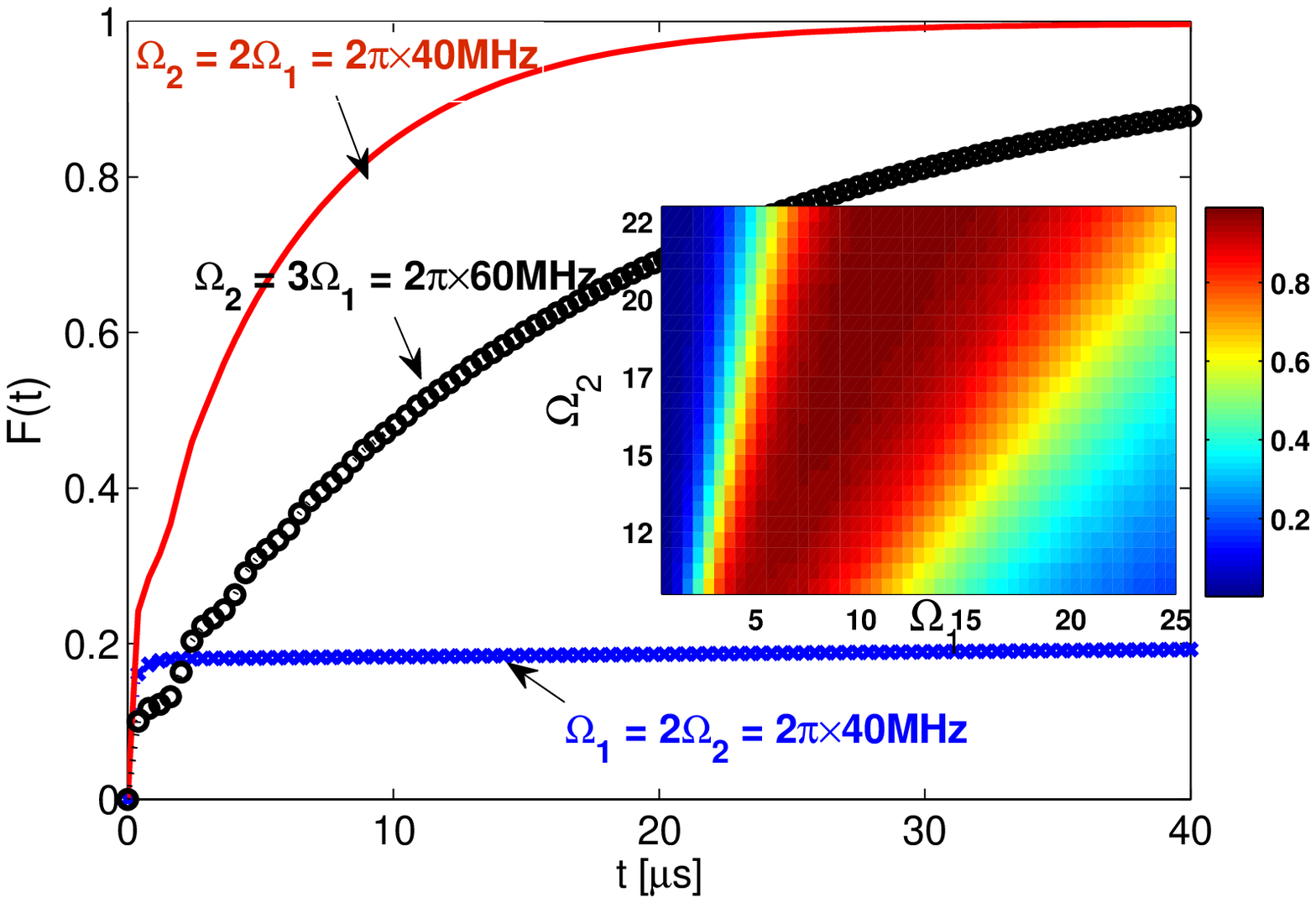}
\includegraphics[width=70mm]{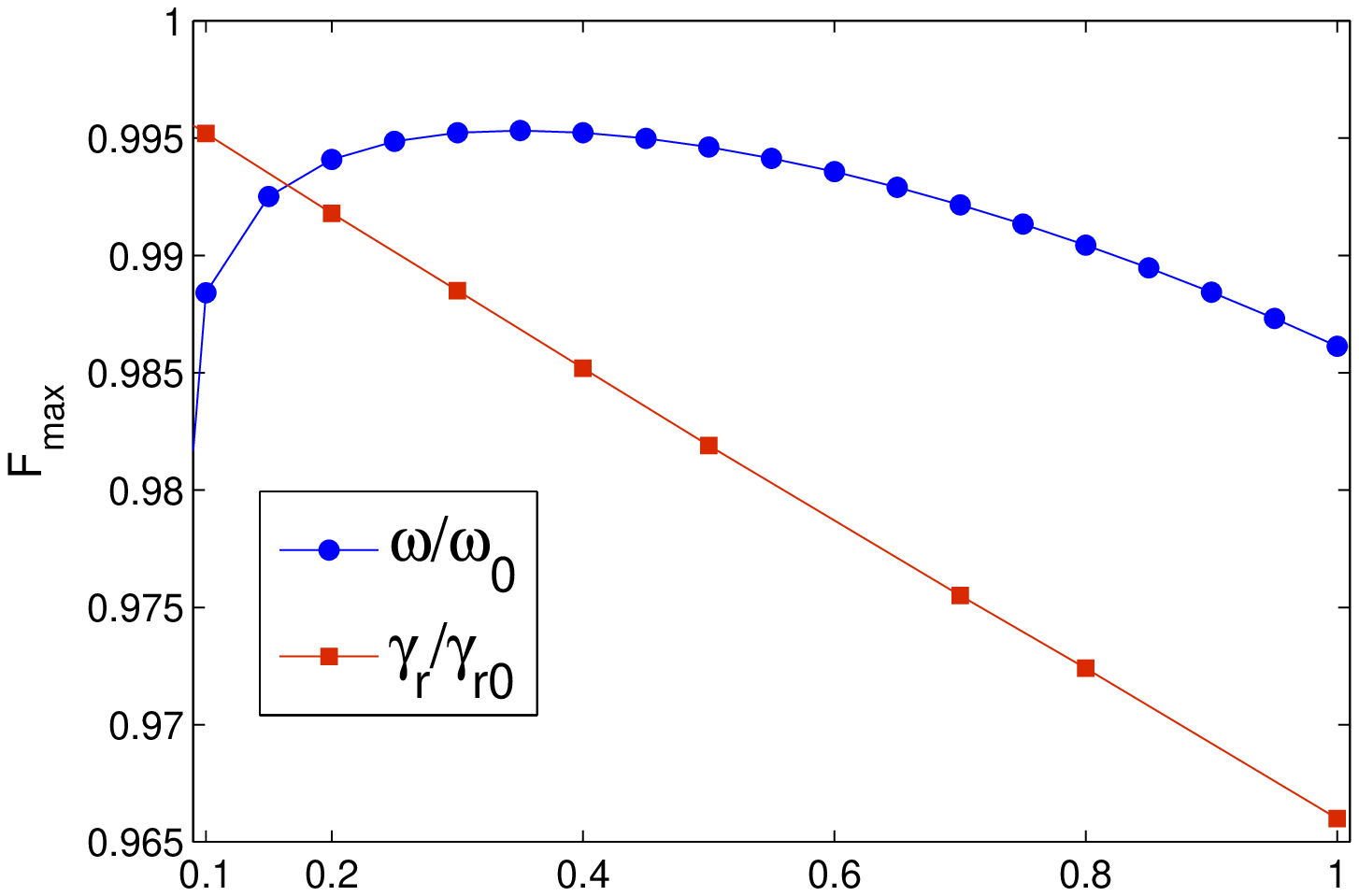}
\caption{(Left) Fidelity is shown as a function of time for three different ratios of the Rabi frequencies. In the inset we have shown $F_{max}$ as function of Rabi frequencies $\Omega_1$, $\Omega_2$. For this plot we have taken $\omega/2\pi = 250$kHz, $\gamma_1/2\pi = \gamma_2/2\pi = 3.03$MHz and $V_{RR}/2\pi = 20$MHz.(Right) The maximum value of fidelity, $F_{max}$ obtained in the long time limit is plotted as a function of $\omega$ and $\gamma_R$. For this plot we have assumed perfect blockade and $\omega_0/2\pi = 1MHz, ~\gamma_{R0}/2\pi = 10kHz$. The values of other parameters used in this plot are similar to those in Figure 1.}
\end{figure*}

\noindent
\subsubsection{Bounds on the rate and magnitude of entanglement generated}
For resonant couplings the Hamiltonian (Eq. 3 main text) has three dark states, $\ket{D_1} = \frac{1}{c_1}\left[\Omega_1\Omega_2\ket{S_{1p}}-(\Omega_2^2-\omega^2)\ket{S_{pr}}+\Omega_1\omega\ket{S_{r0}}\right]$, $\ket{D_2} = \frac{1}{c_2}\left[\Omega_1\ket{S_{1p}}-\Omega_2\ket{S_{pr}}+\omega\ket{S_{10}}\right]$, and $\ket{D_3} = \frac{1}{c_3}[(\Omega^2_2-\Omega^2_1-\omega^2)\ket{11}-\Omega_1\Omega_2\ket{T_{1r}}+\Omega^2_1\ket{pp}+\Omega_1\omega\ket{T_{p0}}-(\Omega_2^2-\omega^2)\ket{00}]$, where $c_i$ are the respective normalization constants and $\ket{S_{ij}} = \frac{1}{\sqrt{2}}[\ket{ij}-\ket{ji}]$ and $ \ket{T_{ij}} = \frac{1}{\sqrt{2}}[\ket{ij}+\ket{ji}]$ are the respective singlet and triplet combinations of the basis vectors $\ket{ij}$, and hence any linear superpositions of the three states also remain dark. In the presence of dissipation neither of these remain dark, as each of them has a nonzero overlap with the decaying level, $\ket{p}$, and hence will gain an imaginary component, allowing any population in these states to decay in the long time limit. While states $\ket{D_3}$ and $\frac{1}{\sqrt{2}}\ket{D_1+D_2}$ attain an imaginary component $\sim \gamma(\Omega_1/\Omega)$ to their eigenvalue the other dark state  $\ket{D_-}=\frac{1}{\sqrt{2}}\ket{D_1-D_2}\approx\frac{1}{\sqrt{2}\sqrt{1+2\xi^2}}[\ket{S_{r0}}-\ket{S_{10}}+3\xi/2(\ket{S_{1p}}-\ket{S_{pr}})]$, where $\xi = \left(\frac{\omega\Omega_1}{\Omega^2}\right)$, gets the smallest imaginary component $\sim 2\xi^2$. Hence $\ket{D_-} \simeq \ket{DS}$ will remain dark even in the presence of dissipation. In fact it is closest to the maximally entangled supersinglet

The bounds on the steady state fidelity and the rate of entanglement can be determined from the eigenspectrum of $\mathcal{H}_{eff}$ (Eq. 5, main text). The maximum fidelity achievable is given by the overlap of the state $\ket{DS}$ with the eigenstate of $H_{eff}$ which has the eigenvalue with the smallest imaginary part. As the imaginary part of the energy determines the decay rate for the corresponding eigenstate the convergence towards the state $\ket{DS}$ is determined by the gap between the two eigenvalues with lowest imaginary component. Assuming perfect blockade the rate of entanglement
\be
\gamma_E \le \omega\frac{\gamma_p\Omega_1}{{\Omega^2_1+\Omega^2_2+\omega^2}},
\ee
and the maximum attainable fidelity is given by
\be
F_{max} \le 1-2\left(\frac{\omega\Omega_1}{\Omega^2}\right)^2.
\ee

\end{document}